\documentstyle[12pt,aaspp4]{article}

\def\etal{{et al.}}
\def\asca{{\it ASCA}}

\def\exosat{{\it EXOSAT}}
\def\ginga{{\it Ginga}}

\def\pcmsq{{$\rm cm^{-2}$}}

\def\nh{{$N_{\rm H}$}}

\newbox\grsign \setbox\grsign=\hbox{$>$} 
\newdimen\grdimen \grdimen=\ht\grsign
\newbox\simlessbox \newbox\simgreatbox \newbox\simpropbox
\setbox\simgreatbox=\hbox{\raise.5ex\hbox{$>$}\llap
     {\lower.5ex\hbox{$\sim$}}}\ht1=\grdimen\dp1=0pt
\setbox\simlessbox=\hbox{\raise.5ex\hbox{$<$}\llap
     {\lower.5ex\hbox{$\sim$}}}\ht2=\grdimen\dp2=0pt
\setbox\simpropbox=\hbox{\raise.5ex\hbox{$\propto$}\llap
     {\lower.5ex\hbox{$\sim$}}}\ht2=\grdimen\dp2=0pt

\begin{document}

\title{ASCA observations of type-2 Seyfert Galaxies: 
III Orientation and X-ray Absorption}

\author {T.J.Turner \altaffilmark{1,2}, 
I.M. George \altaffilmark{1, 2},
K. Nandra \altaffilmark{1, 3},  \nl
R.F. Mushotzky\altaffilmark{1}, 
}

\altaffiltext{1}{Laboratory for High Energy Astrophysics, Code 660,
	NASA/Goddard Space Flight Center,
  	Greenbelt, MD 20771}
\altaffiltext{2}{Universities Space Research Association}
\altaffiltext{3}{NAS/NRC Research Associate}

\slugcomment{Accepted by {\em The Astrophysical Journal}}

\begin{abstract}

We discuss the spectral properties of a sample of type-2 Seyfert galaxies 
based upon the analysis of \asca\ data. 
In this paper we consider the sources for which 
the X-ray spectra appear to be dominated by 
the nuclear continuum, transmitted through a large column of 
absorbing material. 

We find that both Seyfert-2 galaxies and NELGs show iron K$\alpha$ line
profiles indicative of reprocessing of nuclear X-rays in a face-on accretion
disk. Such line profiles are also observed in Seyfert-1 galaxies. This result
is contrary to unification models, which would predict the inner regions of
Seyfert-2 galaxies to be observed edge-on. This raises some questions as to
the orientation of the circumnuclear absorber. If the observed differences
between Seyfert type-1 and type-2 galaxies, and NELGs are not due to
differences in the orientation of the absorbing material, then we suggest that
differences in dust composition and grain size, and in the density of the
circumnuclear gas could be of primary importance.

\end{abstract}


\section{Introduction}

\label{sec:intro}

The classification of an Active Galactic Nucleus (AGN) depends 
upon the wavelength at which one  
observes. Historically, optical observations yielded the categories  
of Seyfert types 1 and 2, classifying most of the 
sources which are now ``famous''.  Seyfert-2 galaxies 
differ from type-1 in that the former show only 
narrow emission lines in their optical spectra. It was soon 
realized that some Seyferts showed weak broad components along with 
the narrow emission lines and consequently 
the subclasses Seyfert-1.5, 1.8 and 1.9 were introduced to 
quantify the differences in strength of the broad-line components 
relative to the narrow lines.
Narrow Emission Line Galaxies (NELGs) are bright and variable 
X-ray sources, discovered in early X-ray sky surveys (Marshall \etal\  1979). 
The narrow optical emission lines 
often have weak, broad H$\alpha$ and P$\beta$ emission 
(Ward \etal\ 1978, Veron \etal\ 1980, Shuder, 1980) making the optical spectra 
similar to Seyfert-1.9 galaxies. Thus the NELG classification is 
indicative that the source was discovered in an X-ray survey, but 
many NELGs are otherwise indistinguishable from Seyfert-1.9 galaxies. 
Optical spectroscopy and spectropolarimetry, infrared spectroscopy, 
X-ray spectroscopy and temporal studies plus $\gamma$-ray spectra are all 
important in the determination of the fundamental nature of obscured nuclei. 

Unification models for AGN postulate that large amounts of dense, molecular 
material exists between the broad-emission-line region (BLR) and 
the narrow-emission-line region (NLR), in some cases within parsecs of  
the active nucleus (see Antonucci 1993 for a review of Unified Models 
for AGN).  
The simplest geometry consistent with observations is a torus, and 
consequently it has been suggested that one of the 
primary factors in Seyfert classification is the orientation 
of the absorbing torus to our line-of-sight. This hypothesis is 
consistent with the existence of circumnuclear molecular 
gas suggested by absorption measurements in a number of wavebands 
(e.g. Braatz \etal\ 1993, Greenhill \etal 1996). 
In unified models, sources observed within the opening angle 
of the torus correspond to those classified optically as 
type-1 AGN, while sources with lines-of-sight intersecting the torus 
correspond to type-2 AGN. In the latter case, the nuclear light can 
nevertheless be 
observed via scattering or transmission. Antonucci \&
Miller (1985) provided compelling support for this model when they detected
broad, Seyfert-1 type emission lines in the polarized optical spectrum of
the Seyfert-2 galaxy NGC 1068, and similar results were later obtained for
a number of Seyfert-2 galaxies (Miller and Goodrich 1990; Tran, Miller and
Kay 1992). Recently, Veilleux \etal\ (1997) provided further support for  
the model with their infrared 
observations of Pa$\beta$, Br$\gamma$ and Br$\alpha$ lines in many Seyfert
2 galaxies, revealing hidden BLRs which were not always detectable in scattered 
optical light. 
Optically-thick torii would be expected to result in collimation of 
nuclear continuum radiation, and 
imaging of optical emission lines has shown preferential elongation of some  
narrow-line regions along the radio axis of the AGN (Haniff, Wilson 
\& Ward 1988). These regions must be ionized by a 
more intense radiation field than is directly observed, again, supporting the 
Unified Model (e.g. Pogge 1988, Tadhunter and Tsvetanov 1989).  
Further support came from \ginga\ X-ray spectra, which showed 
large absorbing columns and iron K$\alpha$ lines of high equivalent width (EW) 
(e.g. Awaki \etal\ 1991). 

The {\it ASCA} satellite (Makishima et al. 1996) consists of four 
co-aligned grazing-incidence X-ray telescopes (XRTs; Serlemitsos 
et al. 1995). The focal-plane instruments are two solid-state 
imaging spectrometers (SISs), each consisting of four CCD chips, 
providing an effective bandpass 
$\sim$0.4--10~keV  (Burke et al. 1994), and two gas imaging spectrometers 
(GISs) at the focus of the other two XRTs,
providing  coverage over $\sim$0.8--10~keV (Ohashi et al. 1996
and references therein).
The sources presented here were systematically analyzed in the same 
way as the broad-line Seyfert galaxies presented 
in Nandra \etal\ 1997 (hereafter N97). 
The analysis method is also described in Paper~I.

In a previous paper we presented 
the basic data-analysis results from a sample of {\it ASCA} observations of 
type-2 Seyfert galaxies 
(Turner \etal\ 1997a, hereafter Paper I), i.e. those having predominantly 
narrow optical emission lines. 
The original sample of 26 observations of 25 
narrow-line AGN comprised 17 Seyfert-2 galaxies and 
8 NELGs drawn from the {\it ASCA} public archive. 
In Paper~I we found the 0.6-10 keV \asca\ spectra of a sample of 
Seyfert-2 galaxies and NELGs to be complex, often containing a 
heavily-absorbed continuum component, a soft X-ray component 
and numerous X-ray emission lines. We found 
the 6-7 keV regime to be dominated by line flux from 
gas with ionization-state $<$ Fe {\sc XVI}. 
Several sites are expected to produce significant X-ray line emission in 
AGN including the line-of-sight absorber, optically-thick material out of the 
line-of-sight (both the putative accretion disk and other larger-scale 
systems such as the torus), ionized (scattering) gas and regions of starburst 
emission. The absence of a strong 6.4 keV iron line component 
in starburst galaxies (Ptak \etal\ 1997), indicates that 
the presence of such a line is likely to be an indication of nuclear activity.  
Iron K$\alpha$ yields the strongest observed X-ray emission line in 
Seyfert galaxies, and thus provides an important probe of conditions 
in the reprocessing material. A discussion of sources 
dominated by scattered and Compton-reflected X-rays 
was presented in a second paper (Turner \etal\ 1997b, hereafter Paper II). 

Fig.~1 (repeated from Paper~II, for ease of reference) shows the equivalent 
width of the iron K-shell line plotted against  neutral 
X-ray absorbing column, $N_H$, for the sample  
sources. Equivalent widths were measured 
against the continuum component dominating the 6 - 8 keV range, and based upon 
the fit to a narrow Gaussian profile (see 
Paper~I for details). The dot-dashed line in Fig~1 denotes the equivalent 
width  of iron K$\alpha$ predicted to be produced by 
transmission through a uniform shell of neutral 
material (with solar abundances subtending 4$\pi$ to a continuum source of 
photon index $\Gamma=2.0$, Leahy \& Creighton 1993), where 
the photon flux $N(E) \propto E^{-\Gamma}$. The dashed line 
shows the equivalent width predicted via Compton-reflection from 
optically-thick material, as a function of \nh, 
assuming that only the power-law is absorbed. In this case the reflection 
is assumed to be produced from the accretion disk with an equivalent width 
$\sim 230$ eV (N97), as typically observed in Seyfert-1 galaxies. 
Coincidently, 230 eV represents both the maximum equivalent width 
observed when iron K$\alpha$ 
was parameterized as a narrow Gaussian line, and the mean equivalent width 
assuming a relativistic line profile (N97). 

Sources can lie significantly above both of these 
model lines if the direct continuum is hidden but the reprocessed emission 
is observed, as the line equivalent width is then measured against 
a suppressed continuum. Consideration of the iron K$\alpha$, 
 \verb+[+O{\sc iii}\verb+]+ $\lambda 5007$ line and X-ray variability 
together suggested that 
NGC~1068, NGC~4945, NGC~2992, Mrk~3, Mrk~463E and Mrk~273 
are dominated by reprocessed X-rays (Paper~II). 
These sources were denoted ``group C'' (marked with squares on Fig.~1). 
Sources lying on the ``Leahy and Creighton line'' were denoted 
``group A'' (marked as circles on Fig.~1). 
Thus group A is composed of NGC~1808, NGC~4507, 
NGC~5252, NGC~6240, ESO~103-G35, IC~5063, NGC~7172 and NGC~7582. 
(Table~1 shows the group designation of the sources 
based upon the original sample). Sources with iron K$\alpha$ equivalent 
widths lying between that line and the 230 eV line 
are consistent with Seyfert-1 spectra transmitted through a high absorbing 
column and are denoted ``group B'' (marked as stars on Fig.~1).  
Group B is composed of NGC~526A, NGC~2110, MCG-5-23-16 and 
NGC~7314, which are all NELGs. This classification implies 
we see the nuclear component directly in group B, and this is 
supported by the observation of rapid X-ray variability in their flux 
(Paper~I and Hayashi \etal\ 1996) and rapid variability of the iron line 
profile in NGC~7314 (Yaqoob \etal\ 1996). 
This division into groups A, B and C leaves 
MCG-01-01-043, NGC~4968, NGC~6251, E~0449, NGC~5135 and 
NGC~1667 unclassified,  i.e. those with 
the lowest signal-to-noise ratio in the \asca\ data. As we will 
demonstrate, this crude classification of sources yields a useful insight 
into the relative importance of the regions contributing to the 
X-ray spectra in several different cases. The distribution of 
sample sources is shown in Table~2. 

To summarize, in Paper~II we showed that in group-C sources the 
iron K$\alpha$ complex contains significant 
contributions from neutral and high-ionization species of iron, thus 
Compton-reflection, hot gas and 
starburst emission all could make significant contributions to the
observed X-ray spectra.  
Mrk~3 appeared to be the only source which had little 
contamination by starburst activity and in this case the {\it ASCA} spectrum 
below 3 keV is dominated by gas 
with an X-ray ionization parameter 
$U_{X}\sim 5$ (as defined by Netzer 1996) and effective column density 
$N_{H} \sim 4 \times 10^{23} {\rm cm}^{-2}$.  This
material is more highly ionized than the zone of material comprising
the warm absorber seen in Seyfert-1 galaxies (George \etal\ 1997, 
hereafter G97), but may
contain a contribution from shock-heated gas associated with the jet. 

In this paper the X-ray properties of sources in groups A and B are examined 
in the context of unified models for AGN. 

\section{Constraints from Emission}

\subsection{The Continuum form}

Paper~I suggested the underlying continuum has a power-law with a slope  
$\Gamma \sim 1.9-2.0$ in the 0.6 -- 10 keV band. 
However, this result might not be meaningful for sources where we used a 
power-law to parameterize a spectrum 
dominated by scattered or reflected components (i.e. group C). 
Taking groups A and B and considering  
the continuum component which dominates the 
2-10 keV band we obtained weighted mean slopes 
$\Gamma_A=1.68\pm 0.06$ and $\Gamma_B=2.11\pm 0.03$ respectively, and 
a dispersion of 0.11 for group A, 0.31 for group B. 

It is difficult to make a strict comparison between our results and 
the indices derived for 
Seyfert-1 galaxies. A slope $\Gamma \sim 2$ was obtained when the latter 
where fit to a model which included the effects of 
Compton-reflection (e.g. G97). Type-2 Seyferts generally have 
complex soft X-ray spectra, and it is difficult to 
constrain the soft component, primary power-law and the effects of 
Compton Reflection, using \asca\ data. However, we can compare the power-law 
parameterization of the 2-10 keV data with a 
similar parameterization of type-1 Seyferts. 
Group A slopes are consistent with the $\Gamma \sim 1.7$ 
slopes obtained when proportional-counter spectra of Seyfert-1 
galaxies were modelled with a  
simple absorbed power-law (Mushotzky 1984, Turner \& Pounds 1989) 
over the $\sim 2-10 $ keV range. Thus group A appear consistent with 
the spectra of Seyfert-1 
galaxies when a comparison is made between similar fits.

In group B, NGC~526A and NGC~7314 show a flattening of the \asca\ 
spectra to high energies (e.g. Table~12 of Paper~I) while the 
indices inferred for the primary continuum  are  
$\Gamma \simeq 1.8$ and $\Gamma \simeq 2.4$ respectively. MCG-5-23-16 
and NGC~2110 appear to be 
dominated by $\Gamma \simeq 1.9$ and 1.6 components. 
The $\Gamma \sim 2.4$ slope is the only one which 
lies outside of 
the range of slopes observed in the Seyfert-1 sample (G97). However, 
the slope implied in the case of NGC~7314 
depends strongly upon whether a hard component 
is allowed. We conclude that 
groups A and B both show spectra which are 
generally consistent with the 
distribution of primary continuum slopes observed in 
Seyfert-1 galaxies.  Comparing 
historical X-ray observations of various sources, it seems 
that some sources show significant index variability, as 
noted by Weaver \etal\ (1997) for MCG-5-23-16. 

\subsection{Iron K$\alpha$ Line Profiles}

As discussed in the introduction, 
most of the sources in the original sample (of 26 observations) 
show a significant 
iron K$\alpha$ line and the mean rest-energy of the line is 
$6.40\pm0.02$ keV (Paper~I). The equivalent widths from the narrow Gaussian 
model are shown in Fig.~1. Consideration of the (weighted mean) 
equivalent widths in the narrow Gaussian model to the line 
yields $113\pm19$ and $97\pm14$ eV for groups A and B 
respectively. However, the narrow Gaussian 
model (with energy fixed at 6.4 keV) 
could miss some of the flux from a broadened line. 
The broad Gaussian model 
yields EWs $153\pm40$ eV and $223\pm45$ eV for A and B. 
The line energies and widths are $E_A=6.37\pm0.05$ keV, 
$\sigma_A=0.21\pm0.07$ keV, 
$E_B=6.33\pm0.05$ keV, $\sigma_B=0.32\pm0.08$ keV. 
To investigate further, we 
constructed average profiles for each group. 

Papers~I and II showed the average SIS+GIS iron K$\alpha$ line profiles for 
the individual sources. However, the SIS data have superior energy resolution to the
GIS data, and so examination of the averaged SIS0+SIS1 data can give the 
clearest insight into the intrinsic line profiles. Obviously, 
consideration of the SIS data alone results in a decrease of 
signal-to-noise ratio when the data are considered on a source-by-source basis.  
However, the SIS data can provide the most useful information  
when the data/model ratios for different sources 
are co-added.  The energy-scale of each SIS spectrum was redshift-corrected 
to the rest-frame of the source. The average SIS data/model ratio 
was then calculated (based upon the best-fit continuum 
model) combining sources within each group. The result 
of this summation is shown for each 
group in Fig.~2.  The dotted Gaussian profile represents the SIS instrument
response for a narrow line observed at a rest-energy of 6.4 keV. The 
profile constructed from the Seyfert-1 sample (N97) is also shown for 
comparison. 

The group C profile shows line flux at and above 6.4 keV 
(a result which is confirmed by consideration of 
the fits to the SIS$+$GIS data for each individual source) 
probably due to 
the presence of unresolved iron K$\alpha$ lines from a range of 
ionization-states (as discussed in Paper~II).   Interestingly,  
the group C profile also shows a red wing. While this is small 
compared to the 
peak of this line, it is of the same amplitude ($\sim 10\%$) 
as that observed in 
Seyfert-1 galaxies, and group A and B. We return to this point  later. 

The group B profile is not significantly distinguishable from that 
observed for Seyfert-1 galaxies (N97). The line peak falls at 
$6.41\pm0.02$  keV, 
and the line core is significantly broadened towards lower energies. 
A very broad wing is also evident, with the broad line component 
peaking at $6.44\pm0.14$ keV, with width $\sigma=0.68\pm0.12$ keV. 

A broad iron line with a red-wing is predicted if the line arises close to the
black-hole and is subject to gravitational redshift. The iron K$\alpha$
emission observed in Seyfert-1 galaxies has been successfully modeled with
profiles expected for a line originating in an accretion disk around a
black-hole (e.g. Tanaka \etal\ 1995, N97), hereafter denoted a disk-line.
Unless the disk is observed face-on, kinematic
(Doppler) effects cause the blue flux of the line to be enhanced
relative to the red-wing, contrary to what is observed. Hence the
requirement for a face-on geometry when the red wing is the dominant
feature observed. In the Seyfert-1 case, 
the iron K$\alpha$ line profile was interpreted as originating in a 
relativistic accretion disk  
inclined close to face-on (the mean inclination of the 
Seyfert-1 sample was $26\pm2\arcdeg$, for a fit where the emissivity 
was fixed as a function of radius; N97, Table~5). 
The similarity in 
line profiles suggests a comparable orientation for the inner regions of 
group B sources, which are all NELGs.

The group A profile has a core which is consistent with 
a narrow Gaussian line, centered at 6.4 keV. 
The equivalent width of the core is 
$113\pm19$ eV which could be produced by 
transmission of the power-law continuum through 
columns of $\sim 10^{23} {\rm cm}^{-2}$ in the line-of-sight (Fig.~1). 
However, this profile also has excess flux in the wings of the line, 
reminiscent of the red-wing observed in group B and in Seyfert-1 galaxies. 
In the light of the group B result, we removed the NELGs from
the group A profile (half the group). 
We found that the Seyfert-2 galaxies alone 
also yield a similar profile with a broad red wing. 

Our finding that group A and B have similar profiles is an 
interesting and surprising result as unification models 
predict the inner regions of type-2 Seyferts to be observed edge-on.

We attempted to fit the iron K$\alpha$ line in individual sources, using 
the disk-line model profile of Fabian \etal\ (1989). That model 
assumes a Schwarzschild geometry and computes the line profile but 
not the line strength. In that model the line emissivity is 
parameterized by a power-law as a function of radius, $r^{-q}$. 
As line and continuum parameters can be correlated, and 
as the fit depends significantly on the assumptions made, we 
adopted the same assumptions as those used to fit a large sample of  
Seyfert-1 galaxies (N97), to allow a meaningful 
comparison between the two sets of results. 

In each case we assumed the line originates from a region between 
6 and 1000 r$_g$, where the gravitational radius, $r_g=GM/c^2$, 
and $M$ is the mass of the black hole. The line emissivity 
function was assumed to be r$^{-2.5}$. We allow the inclination 
of the disk relative to the observer to be a free  
parameter (defined such that $i=0$ is a face-on disk), 
along with the normalization of the line. We assume  the line energy 
${\rm E_{K\alpha}}=6.4$ keV in the rest-frame of the host galaxy. 
The disk-line fits were performed assuming the best continuum models from 
Paper~I with all continuum parameters allowed to be free. 
The continuum models in Paper~I were derived fitting the 
0.6-5.0 plus 7-10 keV data, before consideration of the 
iron K regime was included in the model. In each case we also 
considered the results when an additional (narrow) 
line component was allowed, to represent the iron line produced 
in the line-of-sight absorber. 
The Seyfert-2 galaxies in group A did not generally yield 
useful constraints on the inclination of any accretion disk system. This 
is due to the low signal-to-noise ratio in those sources, and 
the allowance of 
a relatively strong narrow-line component in the fits 
(up to several hundred eV, which could be produced in the line-of-sight 
absorber). 
However, fitting a disk-line profile to the bright sources from group B 
yielded some interesting results.
The rest energy of the narrow line component was fixed at 6.4 keV, 
and the equivalent 
width was constrained to be between 10 -- 50 eV. This range 
embraces the values expected from the Leahy \& Creighton (1993) 
simulations, but allows a wide range about the predicted value, 
to account for differences in spectrum 
for each source, and possible geometric effects. Errors quoted are 1 $\sigma$. 

\subsubsection{Disk-line fits to group B sources}

{\it NGC~526A}

The preferred continuum model for NGC~526A is a double power-law 
with $\Gamma_1=1.82^{+0.16}_{-0.14}$, 
$N_H=1.5\pm 0.14  \times 10^{22} {\rm cm}^{-2}$ 
(Paper~I), flattening to $\Gamma_2=0.78^{+1.07}_{-0.21}$. 
Adding a disk-line profile to the model yields an 
improvement $\Delta \chi^2 =11$ over the fit to a narrow Gaussian line, 
with an inclination 
angle $i=13^{+6}_{-13}\arcdeg$ and EW$=281\pm56$ eV. 
If one assumes instead a single power-law continuum partially-covered 
by the absorber, the fit yields the same inclination and 
consistent EW$=221\pm45$ eV for the line. Addition of a 
narrow Gaussian component (EW $< 50$ eV) did not
significantly improve the fit. 

{\it NGC~2110}

The best-fit continuum from Paper~I was a $\Gamma=1.57\pm0.07$ continuum 
partially-covered by ionized gas with $U_X=0.006^{+0.62}_{-0.005}$, 
$N_{H,ion}=3.9^{+1.71}_{-0.24} \times 10^{22} {\rm cm}^{-2}$. 
The disk-line yielded an 
inclination angle $i=15^{+9}_{-7}\arcdeg$, with 
EW$=289^{+59}_{-52}$ eV and an improvement $\Delta \chi^2 =27$ over the fit 
where the iron K$\alpha$ line was modelled with a narrow 
Gaussian component. Addition of a narrow Gaussian component 
($<50$ eV) did not significantly affect this fit. 

{\it MCG-5-23-16}

The nominal best-fit from Paper~I was $\Gamma_1=1.90\pm0.04$, 
$N_H=1.6^{+0.05}_{-0.07} \times 10^{22} {\rm cm}^{-2}$, 
steepening to $\Gamma_2=4.93^{+0.07}_{-2.08}$ at $\sim 1$ keV. 
A disk-line fit to MCG-5-23-16 yielded a reduction $\Delta \chi^2=66$ (F=60) 
compared to the fit where the 5-7 keV band is modelled with a series of 
narrow lines (Paper~I; Table~15). The line EW was 
$362^{+94}_{-43}$ eV  and the disk inclination was $33^{+11}_{-4}\arcdeg$. 
Addition of a narrow Gaussian component 
($<50$ eV) did not significantly affect this fit. 
Weaver \etal\ (1997) found a somewhat higher inclination angle in their 
analysis of MCG-5-23-16, but all disk-line results depend strongly on the 
assumptions made in the fit, which were different in this analysis versus 
Weaver \etal\ (1997). 
The $i=33\arcdeg$ solution is consistent with the observation of some 
line flux blueward of the peak (in addition to the red wing). 

{\it NGC~7314}

The nominal best-fit to NGC~7314 was $\Gamma_1=2.41^{+0.05}_{-0.05}$, 
$N_H=1.52^{+0.04}_{-0.14} \times 10^{22} {\rm cm}^{-2}$, 
$\Gamma_2=0.97\pm0.26$. The addition of a disk-line component 
yields a good description of the iron K$\alpha$ line profile 
improving the fit by $\Delta \chi^2 = 20$, compared to the 
narrow Gaussian parameterization. The fit yields 
$i=17^{+6}_{-7} \arcdeg$, $EW=417\pm82$ eV. Addition of 
a narrow Gaussian component ($< 50$ eV) did not significantly 
affect this result. 
However, fits to this source are very sensitive to the 
underlying continuum and absorption. Fitting the disk-line on 
top of a single power-law continuum, partially covered by 
a column $8 \times 10^{21} {\rm cm}^{-2}$ produces a 
solution with $i=56^{+34}_{-4}\arcdeg$, $EW=388\pm 76$ eV. 
The addition of a narrow Gaussian component does not significantly affect 
this solution. The former fit is nominally statistically preferred 
(by $\Delta \chi^2 =13$) but obviously the conclusions drawn 
depend strongly on the approach taken in the spectral fitting, this
is the origin of differences in the results obtained here versus 
Yaqoob \etal\ (1996). 

\subsubsection{Consideration of the ubiquity of the red-wing}

We now consider possible explanations for the ubiquity of the  
red wing on the iron K$\alpha$ profiles of Seyfert galaxies. 

Given the correlation between line and continuum parameters, 
an inadequate modelling of the continuum and absorption properties 
might be misleading us about the profile of the iron K$\alpha$ line. 
This is more likely to produce a significant effect in type-2 Seyferts  
than in type-1, since the former show relatively large spectral 
modification due to attenuation in the middle of the \asca\ bandpass. 
However, it would be a strange coincidence for such a problem to so 
closely mimic the profile observed in Seyfert-1 galaxies. 
Consequently we consider this possibility unlikely to be the cause of the 
similarity in observed profiles. 

Another possibility is that the type-2 objects have 
another mechanism for producing a red-wing on the iron K$\alpha$ line. 
The existence of a Compton-shoulder has been suggested for 
sources such as NGC~1068 (Iwasawa \etal\ 1997). 
However, the Compton shoulder cannot easily produce a such a strong wing 
extending down to such low energies as observed in our profiles. 

Despite the systematic effects observed, we note some variety in the 
profiles of individual sources in the Seyfert type-1 and 
type-2 samples. We also note that the dark earth and Coma cluster do not 
show these characteristic line profiles in their \asca\ spectra (N97). These 
facts suggest that the observed profiles are not an artifact of any 
outstanding uncertainties with the \asca\ calibration. 

We also consider the possibility that the iron K$\alpha$ profile is 
not a good indicator of the orientation of the inner regions of the AGN. 
If strong relativistic effects bend the light produced close to the 
central nucleus, then the iron K$\alpha$ profiles could be 
distorted, and our fits to a Schwarzchild profile could be misleading 
as to the orientation of the central source. 
In particular, the Kerr metric allows contributions to line emission 
from a disk extending inwards to within a few gravitational radii, hence 
yielding more severely skewed line profiles than produced around a 
non-rotating black hole (e.g. Laor 1991). 
However, these line profiles also have a distinct dependance on inclination
angle, and as the line profiles of the type-1 and type-2 Seyferts are so 
similar, the preferred orientation of the innermost 
reprocessor must be the same. 
In the discussion we investigate the consequences of this conclusion 
in the context of the unified model.

\section{The effects of obscuration}

Observations at different wavelengths provide probes to 
different depths into the AGN. 
Assuming the circumnuclear absorbing material has a Galactic dust-to-gas 
ratio, then a V band extinction of A$_V=$1~mag. corresponds to a column of 
neutral hydrogen $N_{H,z} \sim 1.5\times 10^{21}\ {\rm cm^{-2}}$. 
An extinction 
A$_V < 5$ allows the broad line region to be observed. A$_V \sim$
5-8 hides the broad optical lines, still these AGN can be
identified by direct observation of the Paschen series of lines in 
the infrared regime. Extinction in excess of A$_V \sim 8$ still 
allow  observations of Pa$\beta$ (which probe material up to 
A$_V \sim 11$), 
Br$\gamma$ (up to A$_V \sim 24$) and Br$\alpha$ (up to A$_V \sim 63$; 
see Vielleux \etal\ 1997). When the extinction approaches A$_V \sim 100$ 
then only X-ray and 
$\gamma$-ray observations are able to provide a {\it direct} view of the
nucleus, and as extinction increases beyond this level, then even the 10 keV
photons are hidden, the material is optically-thick to Compton scattering 
and the central source is only detectable via scattering 
by material out of the line-of-sight. Thus the classification 
assigned to a source may depend on the wavelength at which it is 
observed. Observations in some bandpasses might not show any signature  
of an important component of the system. This has lead to some historical 
confusion as to the nature of some sources.

\subsection{Evidence from ASCA regarding the nature of the the X-ray Absorber 
and Soft Excess}

As evident in Fig.~1, the column densities for group B sources are 
clustered around 
$10^{22}  {\rm cm^{-2}}$  while group A span 
$\sim 3 \times 10^{22} - 10^{24}\ {\rm cm^{-2}}$. As 
discussed in Paper~II, the column densities derived for the scattered X-ray 
sources (group C) generally do not represent the line-of-sight 
absorption to the nucleus. Group~C 
sources were treated in detail in Paper~II and that class 
is not discussed further in this paper.  

Of the sources considered in this paper, only NGC~7172 is 
consistent with a single power-law absorbed by 
neutral material. 
The spectra of all the other sources in groups A and B 
show a strong soft X-ray flux in excess of that predicted by extrapolation of 
the hard X-ray power-law, absorbed by neutral material. A ``soft excess'' 
can arise when there are 
unattenuated lines-of-sight to the nucleus, when the absorber is 
ionized or when a soft emission component is present. 

Of the Group~A sources; NGC~1808, NGC~4507, NGC~6240, ESO~103-G35 
and NGC~7582 have an  intrinsic 0.5 -- 2.0 keV luminosity 
in the soft component (alone) $\sim 10^{40} {\rm erg\ s}^{-1}$, when a 
correction is made only for the Galactic 
line-of-sight column. (We note this yields a lower 
0.5 -- 2.0 keV luminosity than those listed in Tables 8-13 of Paper~I, 
which were corrected for total absorption.) 
Consideration of the maximum starburst contribution to the X-ray flux 
(Paper~I) reveals that only 
NGC~1808 is consistent with all of that soft flux originating from starburst 
regions (Tacconi-Garman, Sternberg \& Eckart 1996), however, 
soft X-ray luminosities $\sim 10^{40} {\rm erg\ s}^{-1}$ could 
be accounted for by the 
summed contributions of starburst emission, 
hot gas and binary stars in the host galaxy (although these 
cannot generally account for the observed hard X-ray luminosities). 
No significant variability is evident in the soft flux or absorbing column in 
NGC~4507, NGC~1808 or NGC~6240 (for which few X-ray observations are 
available), consistent with an active nucleus covered by a uniform absorber 
with an extended soft X-ray emission component providing the spectral 
complexity. 

Variations in  X-ray absorption have been claimed for 
ESO~103-G35 (Warwick \etal\ 1988) on the basis of two \exosat\ observations 
which revealed a drop in column 
$\Delta N_H =8  \times 10^{22} {\rm cm}^{-2}$  in 90 days.  
The timescale for variability indicated that 
absorber to be consistent with clouds at a radius coincident with the BLR. 
The \asca\ observation reveals an absorbing column consistent 
with the most heavily 
absorbed \exosat\ spectrum, as does analysis of a later 
\asca\ observation (26 September 1995), raising the question as to 
whether a contaminating source could have affected 
the apparent measurement of a lower column at one of the \exosat\ epochs. 

The absorption of NGC~7582 also appears lower 
($\Delta N_H \sim 1 \times 10^{23} {\rm cm}^{-2}$) than that 
observed during the \exosat\ observation (Turner \& Pounds 1989). The nearby 
BL Lac PKS2316-423, may have contaminated the \exosat\ 
spectrum obtained with the collimated medium energy (ME) instrument. 
However we would expect any such contamination to have made the
\exosat\ column measurement attributed to NGC~7582 
to appear lower than the \asca\ measurement, 
rather than higher. The nearby clusters Abell 1111S 
and Sersic 159-03 also posed potential sources 
of contamination for the \exosat\ observations, and could have yielded an 
erroneous column measurement.  Examination of the 
\exosat\ archives shows that the source and background observations 
had the ME proportional counter offset and with a roll angle 
designed to minimize the contamination from Sersic 159-03. 
However, given the large number of contaminating sources in the \exosat\ 
ME field, any claim of column variability in NGC~7582 requires confirmation 
using further observations of the source. 

In the cases of NGC~526A, NGC~7314, MCG-5-23-16 and NGC~2110 in group B, 
plus NGC~5252 and IC~5063 from group A, 
the soft emission exceeds that expected from stellar origins by 
at least an order of magnitude. These luminous soft components  
could originate from scattered nuclear flux, or leakage 
of the primary continuum through unattenuated line-of-sight. 
The history of the soft X-ray flux could, in 
principle, distinguish between these two. 
Variations in absorbing column 
could indicate that the absorber had a variable 
ionization-state, covering fraction or column. 
Analysis of a {\it ROSAT} HRI observation of NGC~5252 
(made in 1995 July) showed the soft flux to be a factor of 
several lower than that observed by \asca\ (1994 January). 
As the \asca\ spectrum suggests 
the 0.5-2.0 keV flux is dominated by scattered or 
unattenuated nuclear emission 
this variability favors  partial-covering of the absorber or a 
sub-parsec size for the scattering region.  
Our analysis of the {\it ROSAT} PSPC observation of IC~5063 showed the 
soft flux 
to be consistent with that observed by \asca\, allowing no constraint 
in that case. 
X-ray column variability has been claimed for NGC~2110 on timescales of years
(Hayashi \etal\ 1996) based upon 
X-ray observations covering HEAO-1 (1978) to \asca\ (1994), although we note 
that it is difficult to make a conclusive comparison based upon different 
spectral fits performed across a variety of bandpasses, and this result needs
confirmation. NGC~526A showed strong spectral changes, 
consistent with column variability, on a 
timescale of years across five \exosat\ observations (Turner \& Pounds 1989). 
Historic X-ray observations of NGC~7314 
and MCG-5-23-16 show column measurements consistent with no variation.

A power-law source partially-covered by neutral material would be expected to
show a leaking fraction with the same spectrum as that observed in the hard
X-ray regime.  Simple electron scattering of the  hard continuum is also
expected to yield the same spectrum in scattered X-rays, as the primary
continuum. However, many  datasets marginally prefer the double power-law model
(Paper~I) with some sources showing a steep soft X-ray component, and others a
hardening of the spectrum to higher energies. This indicates some spectral 
curvature which is not accounted for by the aforementioned models. 
This could be due to curvature in the underlying continuum. 
However, partial-covering by ionized material could also yield a difference in 
soft and hard spectra. 
This is the preferred model for Seyfert~1 galaxies (Reynolds 1997, G97) 
and for NGC~2110 and NGC~5252 from the type-2 sample. Thus we  
consider the model in more depth in the discussion.

In summary, there is only 
tentative evidence for column variability in the sample sources. The 
the spectral shapes can generally be described either by 
a continuum attenuated by a 
patchy or ionized absorber, or by the presence of a second 
emission component. X-ray monitoring of one or more 
well-chosen type-2 Seyferts will help clarify the picture.

\section{Discussion}

\subsection{Unification with Seyfert-1 Galaxies - Orientation}

Both of our subsamples of type-2 Seyferts show an iron 
K$\alpha$ profile which contains a significant 
red-wing at the same level as that observed in the Seyfert-1 sample. 
This skewed profile implies that type-2 
Seyferts contain an iron K$\alpha$ line component from 
a reprocessor close to the central black hole. 
Furthermore this reprocessor is implied to be 
orientated preferably face-on to our line-of-sight.

This result is contrary to unification models, which 
predict the inner regions of type-2 Seyfert galaxies are 
observed preferentially edge-on. This is all 
particularly interesting in the light of 
the fact that the host galaxies of type-2 Seyferts are also observed 
preferentially face-on. Specifically, Kirhakos \& Steiner (1990) find 
the distribution of inclinations to be indistinguishable between 
type-1 and type-2 hosts when they consider a sample of 288 
Seyfert galaxies. Furthermore, both Seyfert type-1 and type-2 samples are 
observed preferentially face-on compared to a large control sample of normal 
galaxies.  Even NGC~1068 and Mrk~3, which appear to be observed in 
scattered and reflected X-rays, are 
decidedly face-on (with b/a=0.89 and 0.72, respectively). 
However, Maiolino \& Reike (1995) made a study where the ``intermediate'' 
Seyfert 1.8 -- 1.9 galaxies were treated separately to the 
classical Seyfert~2 galaxies. 
They found that both Seyfert~1 and Seyfert~2 galaxies are preferentially 
face-on, but the ``intermediate''  Seyferts are preferentially edge-on. 
This result inferred that the absorption in the Seyfert~1.8 -- 1.9 
galaxies might often be due to material in the plane of the host galaxy 
rather than circumnuclear material. 
NELGs are ``intermediate'' Seyferts on the basis of their optical 
properties, 
however, as the mean iron K$\alpha$ profiles clearly indicate a preference for 
a face-on orientation of the inner regions, there  
cannot be many edge-on sources in our sample or the mean 
iron K$\alpha$ profiles would not be so similar to the mean profile 
for type-1 Seyferts. 

If the putative accretion disk is totally decoupled from the 
orientation of the host galaxy, then we should observe a 
variety of inclinations in the iron K$\alpha$ profiles, 
resulting in a mean profile of $60 \arcdeg$ for the sample.
Thus the preference for a face-on orientation of the nuclear regions 
and the host might suggest a link between the angular momentum of the two. 
This might be expected if the accretion process ultimately 
feeds on material lying in the plane of the galaxy. If this is true, then 
it is difficult to understand why the putative 
obscuring torus, which (if it is present) must lie between these two systems, 
would be 
decoupled from this preferred orientation. This problem might be avoided 
if the absorber is a distribution of clouds, and the 
type-1 or type-2 nature of a source is simply determined by whether or not 
a cloud lay across the observers line-of-sight. Differences in the distribution 
and composition of the clouds could then account for differences in observed 
properties.  Several Seyfert galaxies vary back and forth between 
type-1 and type-2 properties on timescales of years. 
In at least some sources the 
observed behaviour has been attributed to variations 
in the reddening, consistent with clouds of dust moving 
across the line-of-sight (e.g. Goodrich 1989). 
The timescales and repetition of this type of behaviour 
argues against both evolutionary 
models and orientation of large-scale  structures of the absorbing material 
to explain the difference between  Seyfert classes, although the 
``orientation effect'' of a cloud being in the line of sight is 
consistent with the observations. 

In the context of cloud models, another possibility 
is that all classes of Seyfert-type AGN show a similar 
mean iron K$\alpha$ profile because the line is 
produced by reprocessing in a spherically symmetric distribution of 
material. Spherical accretion of clouds of material 
could produce a line showing the strong gravitational effects 
observed, but with the same profile observed from each viewing direction. 
The construction of a mean profile for  a sample of edge-on Seyferts would 
test this theory. The intermediate Seyfert galaxies might 
provide an edge-on sample. Alternatively, 
Kirhakos \& Steiner (1990) find that IRAS galaxies are observed 
preferentially edge-on.  
If the orientation of the host galaxy is 
associated with that of the inner accretion disk/reprocessing region 
then IRAS galaxies should show an iron K$\alpha$ profile indicative of 
edge-on systems, although such a line will be broad, and possibly 
difficult to detect. 

We note that 
the edge-on galaxy NGC~2992 (b/a=0.27) shows an iron K$\alpha$ line 
profile with a blue wing (Paper~I, Paper~2). Weaver \etal\ (1996) 
suggest that the large equivalent width of the iron K$\alpha$ line in 
NGC~2992 is due to a lag of several years between the direct and reprocessed 
components. The data are consistent with a narrow component 
of high equivalent width, as suggested by Weaver \etal\ (1996), plus 
a disk-line component from an edge-on accretion disk, which would support 
a connection between the orientation of the host galaxy and the inner regime. 
However, the data are also consistent with contributions to the iron K$\alpha$ 
line profile from highly ionized species of iron.  A search for  
variability could distinguish between these possibilities, as the 
occurrence of rapid variability in the blue wing would favor the 
disk-line scenario. 

The existence of ``ionization cones'' in a dozen or so Seyfert-2 galaxies 
(e.g. Pogge 1988, Tadhunter and Tsvetanov 1989) 
poses an interesting problem if we assume the 
inner regions are viewed with a face-on orientation. 
Such cones should only be seen in systems observed  
at a different angle to the escape direction of the radiation. 
However, many ionization cones are one-sided. This would be expected if 
we observe close to the direction of the collimated radiation, as 
one side of the 
cone will escape in a direction almost directly away from us. In this case 
the cone will be foreshortened and the back-side will be 
heavily absorbed by any intervening galactic material. We note that 
one of the 
best examples of a two-sided cone is in NGC~5252, in which the host galaxy 
is observed at an intermediate orientation (b/a=0.37).

\subsection{Unification and  Absorption}

The fact that observations of the iron K$\alpha$ line profiles imply a 
near face-on geometry for the inner regions of both Seyfert types 1 and 2 
necessitates a discussion of ways in which factors 
other than the orientation of the absorbing material can explain the 
observed differences between the two classes.

The obscuration of the NLR is similar in optically selected Seyfert-1 and
Seyfert-2 galaxies, indicating that the obscuration responsible for the
differences between these classes is located within the NLR, or out of the
line-of-sight to the NLR (Lawrence \& Elvis 1982, De Zotti \& Gaskell 1985).
In NELGs the X-ray absorbing columns are $10^{22} - 10^{23} {\rm cm}^{-2}$ 
(NGC~7314 to ESO~103-G35) while the optical lines imply an attenuation 
$< 7 \times 10^{21} {\rm cm}^{-2}$ (assuming a Galactic gas-to-dust ratio). 
This implies that the X-ray 
absorber lies within the BLR, or is highly ionized, 
otherwise it would produce stronger suppression 
of the BLR than that which is observed. 

In Seyfert-2 galaxies X-ray column densities are $10^{23}\ {\rm cm}^{-2}$ or
greater. Extinction of the broad lines occurs for columns in excess of $1
\times 10^{22} {\rm cm}^{-2}$. On this basis, the X-ray absorption in
Seyfert-2 galaxies can only be said to exceed that required to fully suppress
the broad optical lines, yielding no constraint on the location of the 
absorber. 

Granato, Danese and Franceschini (1997) 
compare the infrared broad-band 
spectra and spectroscopy results with \ginga\ X-ray spectra of 
a sample of type-2 Seyferts (Smith \& Done 1996). 
For convenience we 
define the infrared extinction as $\tau_{IR}$ and the X-ray 
extinction as $\tau_X$ where 
$I_{obs} = I_{int} e^{-\tau}$, $I_{obs}$ is the observed flux, and 
$I_{int}$ is the intrinsic flux. 
Granato \etal\ find the infrared extinction, $\tau_{IR}$ 
is approximately consistent with the 
X-ray extinction, $\tau_X$, for columns $\lesssim 10^{22} {\rm cm}^{-2}$, 
while $\tau_{IR} \sim 0.1 \tau_X$ for 
$3 \times 10^{22} < N_H < 10^{24} {\rm cm}^{-2}$,  and 
$\tau_{IR} << 0.1 \tau_X$ for $N_H > 10^{24} {\rm cm}^{-2}$. 
Thus, their finding indicates that where there is a large 
X-ray absorption, the material likely 
sits predominantly within the dust sublimation 
radius. 
 
\asca\ evidence for an ionized absorber covering 
NGC~2110 and NGC~5252 makes it interesting to 
examine the general results from such a  model, 
particularly in the light of the strong evidence 
for ionized absorbers in most Seyfert-1 galaxies 
(e.g. Reynolds 1997, G97). 

To examine the unification of type-1 and type-2 Seyfert galaxies we 
examined the distribution of column, ionization parameter 
and luminosity for both types. We took the Seyfert-1 
results from G97. Those sources were analyzed in the same way 
as our sample of type-2 Seyferts, allowing a direct comparison to be 
made. Fig.~3 shows $U_X$ and $N_H$, the ionization-state and 
column of ionized material versus (unabsorbed) X-ray luminosity 
adapted from a figure in G97 (group B sources are 
marked with stars, group A with circles, Seyfert-1 galaxies with 
open squares). 
This comparison demonstrates that while the mean column appears to be 
higher for type-2, some Seyfert-1 galaxies are also 
consistent with having 
large columns of circumnuclear material (which is so highly ionized that 
it does not provide much opacity below $\sim 3$ keV).  

Examination of Fig.~3 shows that there is a large overlap in intrinsic 
(absorption-corrected) luminosity of the type-2 and type-1 sample sources, 
although the mean intrinsic luminosity of  the type-2 Seyferts is lower. 
For gas of the same density and distance from the ionizing source, then 
the ionization parameter is, by definition, proportional to the ionizing 
luminosity. The Seyfert~1 sample appears consistent with 
$U_X \propto L_X$, and hence the circumnuclear gas 
is consistent with similar density and location 
across the type-1 sample.  
However, type-2 Seyferts are clearly inconsistent with that picture, 
most notably 
due to several sources absorbed by material of very low ionization-state. 
We also note that overlaying three quasars on this plot 
(marked with solid squares; 
from George \etal\ 1997b) indicates that there is no 
simple proportionality between $U_X$ and $L_X$.  
This indicates that the absorbing material in 
type-2 Seyferts exists under significantly different conditions than that 
in type-1 galaxies. Differences in gas density, 
geometry and location between individual objects 
could cause the observed effects. 
Type-1 Seyferts do not show such 
pronounced suppression of the broad optical lines 
as type-2 sources with the  
same nuclear (X-ray) luminosity. 
This indicates systematic differences in the location of the dust sublimation 
radius between classes of AGN. 

Reverberation studies have indicated BLR sizes 
$r_{BLR} \sim 0.02 L_{45}^{0.5}$pc ( Peterson \etal\ 1993; 
Netzer 1990), although the BLR probably extends over 
an order of magnitude in radius (Korista \etal\ 1995). 
For bolometric luminosities $10^{43} - 10^{44} {\rm erg\ s}^{-1}$  
(appropriate for our sample sources) we estimate the BLR to be 
$6 \times 10^{15} - 2 \times 10^{16}$ cm ($\sim$ a few light days). 
Barvainis (1987) shows that the dust 
sublimation radius, $r_{sub}=0.13 L_{UV44}^{0.5} T_{1500}^{-2.8}$ pc, where 
$L_{UV}$ is the UV luminosity in units of $10^{46} {\rm erg\ s}^{-1}$ and 
$T_{1500}$ is the grain sublimation temperature in units of 1500 K. 
Order-of-magnitude estimates show the dust sublimation radius 
to lie close to or just outside of the BLR. 
Barvainis also notes that the highest sublimation temperature occurs 
for dust composed of graphite grains, and is $\sim 1500$ K. Dust containing 
silicate grains would have a lower temperature and hence a 
larger sublimation radius. Evidently, different mixtures of dust 
grains will allow the sublimation radius to change by up to a factors of a few. 
The size of the dust grains yields an  
even bigger effect. Laor \& Draine (1993) show that graphite grains 
of size $\sim 10\ \mu m$ can exist almost an order of magnitude closer to 
the nucleus than grains of size $\sim 0.005\ \mu m$. Thus differences 
in the distribution of grain sizes could allow significant 
differences in the location of $r_{sub}$ relative to the 
BLR in different sources, $r_{sub}$ will be closer to the nuclei in 
those sources with the largest dust grains.  However, 
as also shown by Laor \& Draine (1993) 
some distributions of grain size yield dust which is ineffective 
at reddening, i.e. the effective extinction can have a flat dependance 
over a large frequency range. 
The data require that for Seyfert-2 galaxies, the BLR is  
deeply situated well within $r_{sub}$, for NELGs the BLR 
is probably very close to $r_{sub}$ 
because the broad lines suffer some obscuration. Thus differences in dust 
content and in the distribution of grain sizes could be the key to 
unification of type-1 and type-2 Seyfert galaxies. In fact 
Konigl \& Kartje (1994) suggested a dusty wind model to unify AGN, 
with partially-ionized gas in the inner regions of the wind 
accounting for the bulk of X-ray absorption seen in face-on sources. 

Spectral variability in the optical and X-ray regimes could be driven by 
changes in the nuclear flux, which will move the 
effective radii of ionization and sublimation. 
Differences in the distribution of circumnuclear material 
and in the total mass of gas surrounding the nuclei 
may also play a role. A model where clouds of optically thick material 
cause the X-ray obscuration can explain many observed 
properties (Guilbert \& Rees 1988). 
The location, size and column density of the clouds can 
determine whether or not column variability or a leaking fraction of nuclear 
flux are 
observed. Compton-reflection can occur from the surface of the blobs 
(Nandra \& George 1994) 
while scattering of nuclear radiation can occur from a region of hot gas, 
so long as 
there are some unattenuated lines of sight by which to view this. 
Collimation of radiation (to produce ionization cones) could also occur, if the 
blobs are distributed around a preferred plane. 

Interestingly, Maiolino \etal\ (1995) find differences in infrared 
colors indicating that the host 
galaxies of type-2 nuclei have significantly higher levels of 
starburst activity than type-1 nuclei. This starburst activity occurs 
close to the nucleus, and as starburst regions contain 
large amounts of dusty gas, perhaps 
the presence of these starburst regions results in a different 
gas and dust content in type-2 Seyfert galaxies. 
Thus we concur with the suggestion of 
Maiolino \etal\ that consideration of the starburst regions 
seems likely to provide an important component of unification models. 
Differences in dust content of the circumnuclear gas might also 
be related to the apparent preferential occurrence of masers 
in type-2 Seyferts. 
The relationship between a high 
X-ray absorbing column and the existence of water masers in type-2 
Seyferts  
(Braatz, Wilson \& Henkel 1997) indicates that the material in which 
the maser amplification occurs is related to significant X-ray opacity. 
{\it If} the molecular gas is in a well-defined plane then  
masers will be preferentially observed in sources viewed within 
a few degrees of that plane.  
The sources in our original sample which are known to 
have masers are NGC~1068, NGC~4945 and ESO103-G35 (Braatz, Wilson \& 
Henkel 1997). NGC~1068 and NGC~4945 
are so highly absorbed that the X-ray spectra are seen only in reprocessed 
photons, and the \asca\ data are consistent with the high degree of 
opacity expected from an edge-on masing disk. This again raises the 
interesting question as to the relationship of any preferred planes 
of material on size scales of the galaxy, the torus/masing disk or clouds, 
and the nucleus. 
The fact that masers are only observed in type-2 Seyferts might 
indicate that conditions in those AGN  result in a layer of shielded 
masing material. Kartje, Konigl \& Elitzur (1996) propose 
that maser flux arises from clouds in a magnetized accretion disk 
and that these clouds are shielded from the nuclear flux by gas and 
dust. They suggest the inner edge-of the maser is dictated by the 
dust sublimation radius, where effective shielding begins.

\section{Conclusions} 

Examination of the iron K$\alpha$ line from a sample of NELGs and
Seyfert-2 galaxies shows profiles 
similar to those observed in Seyfert-1 galaxies, and 
indicative of an origin in an accretion disk orientated
face-on. This result is contrary to unification models, which 
predict the inner regions of Seyfert-2 galaxies to be observed edge-on. 
The preference for 
a face-on orientation of both the accretion disk and host galaxy of type-1 and
type-2 Seyfert galaxies poses some questions as to the orientation and 
geometry of gas comprising the circumnuclear 
absorbing material. If the absorber is composed of clouds 
then differences in gas density, size of dust grains and dust composition 
and the distribution of the clouds seem a likely explanation of 
the observed differences 
in the X-ray and optical absorption properties of Seyfert-1 galaxies,
Seyfert-2 galaxies and NELGs. These results indicate that further refinement 
is required for unified models for AGN.

\section{Acknowledgements}
We are grateful to \asca\ team for their operation of the satellite 
and to Keith Gendreau, Tahir Yaqoob and Ski Antonucci 
for useful comments. This research has 
made use of the NASA/IPAC Extragalactic database,
which is operated by the Jet Propulsion Laboratory, Caltech, under
contract with NASA; of the Simbad database, 
operated at CDS, Strasbourg, France; and data obtained through the 
HEASARC on-line service, provided by NASA/GSFC. We acknowledge the 
financial support 
of Universities Space Research Association (IMG, TJT) and 
the National Research Council (KN).

\newpage

{\bf Figure Captions}

Fig 1 - Equivalent width of the narrow 6.4 keV line 
versus absorbing column (from Paper~I) for the full sample  
(excluding NGC~1667 which had a line equivalent width of zero, and 
NGC~5695 which was not detected).  The dot-dash line shows 
the line equivalent 
width expected from a uniform shell of material encompassing the
continuum source (Leahy \& Creighton 1993). The dashed line  shows the
predicted equivalent width from reflection as a function of \nh,
assuming that only the power-law is absorbed, but that the reflection
component remains unchanged. Data points are annotated with an
abbreviation of the source name. Group A sources are marked as circles, 
group B sources as stars and group C sources as squares. 
A few of the lowest signal-to-noise datasets have not been classified. 

Fig 2 - The average data/model ratios 
showing the  iron K$\alpha$ line emission for each group, 
and that of the Seyfert-1 sample from N97, for comparison. 
Each dataset was 
corrected to the rest-frame energy before the objects were combined 
and so the  X-axis shows the rest-frame energy. 
The dotted Gaussian profile represents the SIS instrument 
response for a narrow line observed at a rest-energy of 6.4 keV. 

Fig 3 - The absorption-corrected 2--10 keV luminosity versus 
ionization-state and column of the circumnuclear gas, $U_X$ and 
$N_H$. The open squares represent the Seyfert-1 sample of George 
\etal\ (1997), the open circles are the group A sources, stars are 
group B and filled squares represent three quasars (see \S4.2). 

\clearpage
\pagestyle{empty}
\setcounter{figure}{0}
\begin{figure}
\plotone{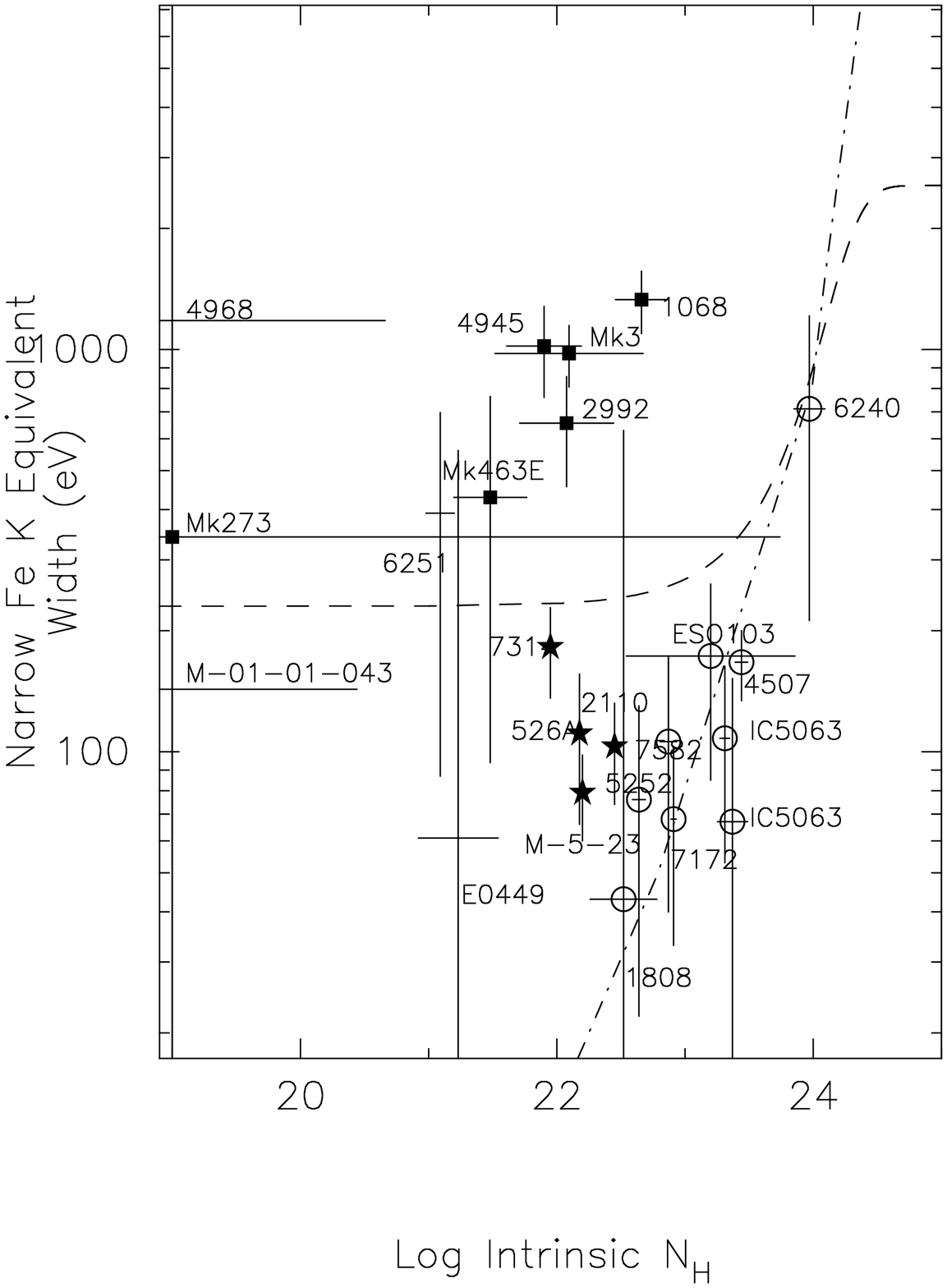}
\caption{}
\label{fig:1}
\end{figure}
\clearpage

\setcounter{figure}{1}
\begin{figure}
\plotone{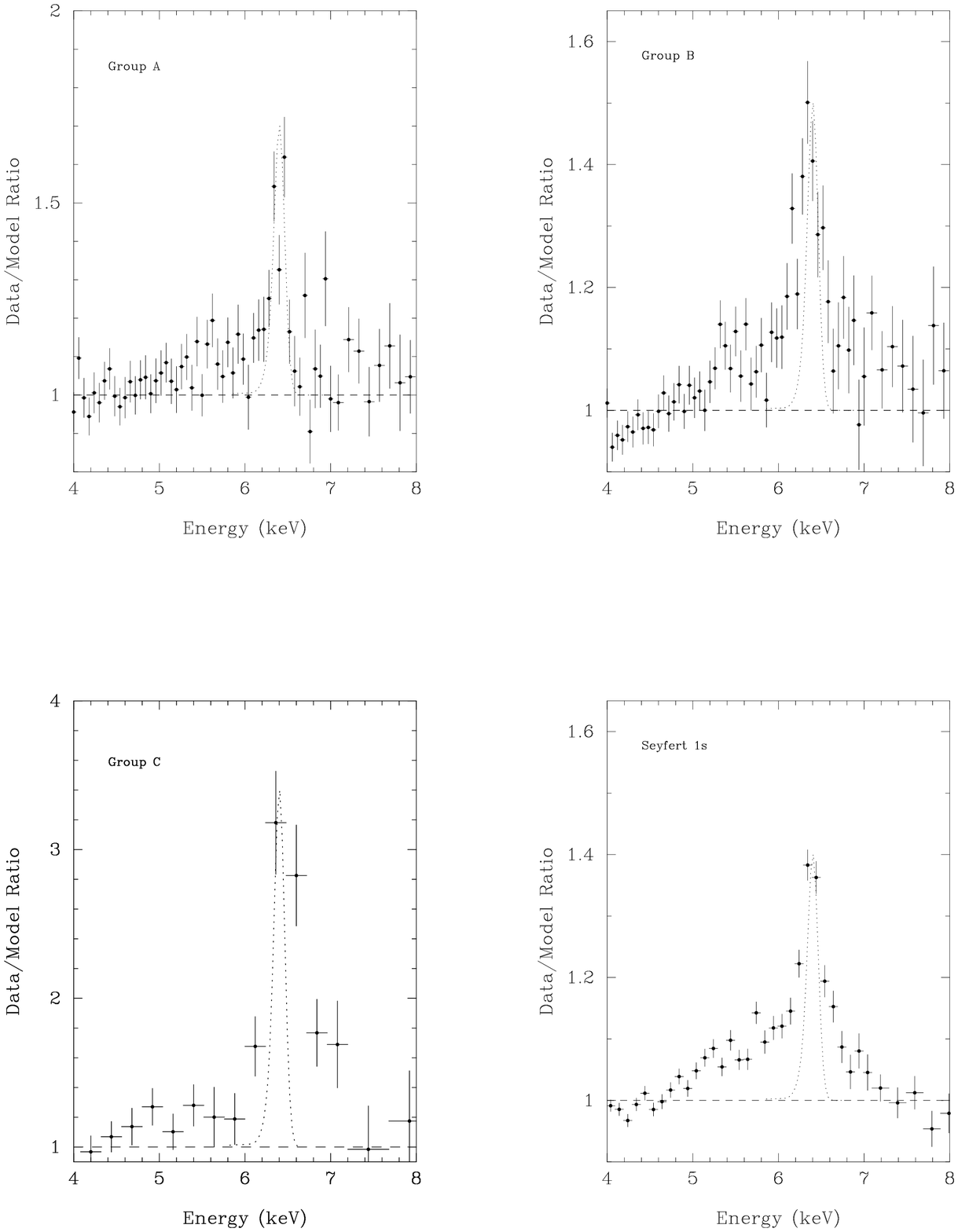}
\caption{}
\label{fig:2}
\end{figure}

\setcounter{figure}{2}
\begin{figure}
\plotone{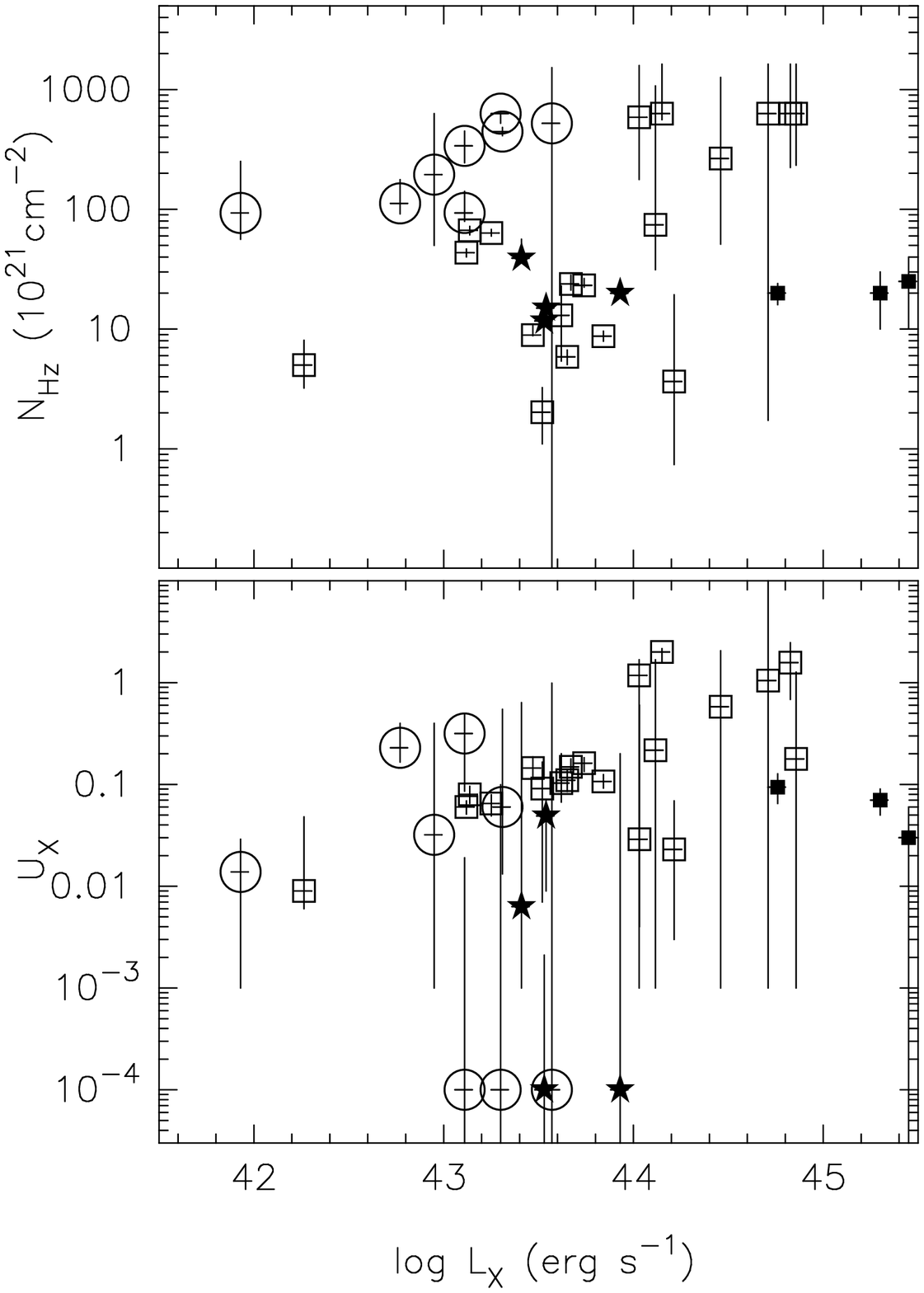}
\caption{}
\label{fig:3}
\end{figure}

\clearpage
\setlength{\topmargin}{-0.5in} 

\begin{deluxetable}{l l c c c c c}

\tablecaption{The \asca\ Seyfert-2 sample. \label{tab:sample}}

\tablehead{
\colhead{Name} & \colhead{b/a} & \colhead{RA}$^{a}$
&  \colhead{DEC}$^{a}$ &
\colhead{z$^a$} & \colhead{Class$^a$} & \colhead{\nh (Gal)} \\
}

\startdata
MCG-01-01-043 &1.0 &00 10 03.5& -04 42 18&0.0300 & S2 & 3.27$^{b}$ \nl
NGC 526A (B)&0.70&01 23 55.1&-35 04 04 &0.0192&NELG$^e$&2.33$^{b}$\nl
NGC 1068 (C)  & 0.89  &02 42 40.8&-00 00 47&0.0038&S2$^{e,h}$& 2.93$^{c}$ \nl
NGC 1667 & 0.83& 04 48 37.2  &-06 19 12& 0.0152 & S2 & 5.46$^{c}$\nl
E0449-184 &\nodata&04 51 38.8 &-18 18 55 & 0.338 & S2 & 3.73$^{c}$ \nl
NGC 1808 (A) & 0.60 & 05 07 42.3 &-37 30 46 & 0.0033 & SB/S2 &2.42$^{c}$ \nl
NGC 2110 (B) & 0.79  &05 52 11.4 &-07 27 22& 0.0076& NELG &  18.60$^{b}$ \nl
Mkn 3 (C) &0.72  & 06 15 36.3 &71 02 15 & 0.0135& S2$^e$&8.70$^{c}$ \nl
NGC 2992 (C) & 0.27& 09 45 41.9&-14 19 35&0.0077&NELG$^e$&5.56$^{b}$ \nl
MCG-5-23-16 (B)&0.45&09 47 40.2&-30 56 54&0.0083&NELG$^e$& 8.82$^{c}$ \nl
NGC 4507 (A)& 0.80 & 12 35 36.5 & -39 54 31& 0.0118 & S2 &7.04$^{c}$ \nl
NGC 4945 (C)&0.13 &13 05 26.2 &-49 28 16&0.0019 & S2$^h$ &  \nodata \nl
NGC 4968  & 0.47 &13 07 06 & -23 40 43 & 0.0100 & S2 & 8.37$^{c}$ \nl
NGC 5135 & 0.80 &13 25 44 &-29 50 01 &0.0137 & S2 & 4.66$^{c}$ \nl
NGC 5252 (A)&0.37 &13 38 15.9 &04 32 33&0.0230&S1.9&1.97$^{d}$ \nl
Mkn 273 (C) & 0.27 &13 44 42.1 & 55 53 13 &0.0378 & S2 & 1.02$^{b}$ \nl
Mkn 463E (C) & 0.57 &13 56 02.9 &18 22 19 &0.0500 & S2$^e$&2.08$^{c}$ \nl
NGC 5695   &0.73 &14 37 22.0 &  36 34 04& 0.0141 &S2 & 1.01$^{c}$ \nl
NGC 6251   & 0.97  &16 32 31.9 &82 32 17 &0.0230 & S2 & 5.48$^{c}$ \nl
NGC 6240-49 (A)& 0.43 & 16 52 59.3 &02 23 59& 0.0245&S2& 5.45$^{c}$ \nl
ESO 103-G35 (A) & 0.51 &18 38 20.2 &-65 25 42& 0.0133&S2/NELG$^h$&\nodata \nl
IC 5063 (A)&0.67&20 52 02.9 & -57 04 14& 0.0113 & S2$^f$ & \nodata  \nl
NGC 7172 (A)&0.42 &22 02 02.1 & -31 52 12&0.0086&S2/NELG$^e$&1.63$^{b}$ \nl
NGC 7314 (B)&0.43&22 35 45.7 &-26 03 03 &0.0047 & S1.9/NELG & 1.45$^{b}$ \nl
NGC 7582 (A) &0.46&23 18 23.2& -42 22 11&0.0053&S2/NELG$^g$&1.47$^{b}$ \nl
\tablenotetext{a}{J2000, from the NASA Extragalactic Database (NED)}
\tablenotetext{b,c,d}{Galactic H{\sc i} column density
in units of $10^{20}$~\pcmsq, Stark \etal\ (1992), HEASARC or 
(c) Elvis, Lockman \& Wilkes (1989), (d) Dickey \& Lockman (1990)}
\tablenotetext{e,f,g,h}{Polarized broad lines detected: (e) Tran 1995,
(f) Inglis \etal\ 1993, (g) Heisler, Lumstron, Bailey 1997; (h) Water 
maser detected; Braatz, Wilson \& Henkel, 1997}
\tablecomments{Column 1: Name (Group);  Column 2: Inclination, from NED; 
Columns 3 \& 4: RA \& DEC; Column 5: Redshift; Column 6: 
Seyfert type, SB= starburst galaxy, NELG=Narrow-Emission-Line Galaxy; 
Column 7: Galactic $N_H$}

\enddata
\end{deluxetable}

\clearpage
\newpage

\begin{deluxetable}{l c c c c}

\tablecaption{Population of the groups}

\tablehead{
\colhead{Type} & \colhead{A} & \colhead{B} 
&  \colhead{C} &  \colhead{Unclassified} 
}
\startdata
NELG/S1.9$^a$ & 4 & 4 & 1 & \nodata \nl
Seyfert-2 & 4 & \nodata & 5 & 7 \nl
Seyfert-1$^b$ & \nodata & 18 & \nodata & \nodata \nl

\tablenotetext{a}{Objects with both NELG and S2 classifications in the 
literature are designated as NELGs here} 
\tablenotetext{b}{N97, this entry included as a reminder that the B class 
is defined by the properties of Seyfert-1 galaxies}

\enddata
\end{deluxetable}

\end{document}